\documentclass[useAMS,usenatbib]{mn2e}
\usepackage{natbib}
\usepackage{graphicx}
\usepackage{amsmath}
\usepackage{amsfonts}

\title[Nonparametric Transient Classification using Adaptive Wavelets]{Nonparametric Transient Classification using Adaptive Wavelets}
\author[M. Varughese, R. von Sachs, M. Stephanou and B. Bassett]{Melvin M. Varughese$^{1}$\thanks{E-mail:
melvin.varughese@gmail.com}, Rainer von Sachs$^{2}$, Michael Stephanou$^{1}$ and Bruce A. Bassett$^{3,4,5}$\\
$^{1}$Department of Statistical Sciences, University of Cape Town, Cape Town, South Africa\\
$^{2}$Institut de Statistique, Biostatistique et Sciences Actuarielles, Universit\'e catholique de Louvain, Louvain-la-Neuve, Belgium\\
$^{3}$African Institute of Mathematical Studies, Muizenberg, South Africa \\
$^{4}$South African Astronomical Observatory, Observatory, South Africa \\
$^{5}$Department of Mathematics and Applied Mathematics, University of Cape Town, Cape Town, South Africa.}
\begin{document}

\date{Accepted XXXXXXXXXX. Received XXXXXXXXXX; in original form XXXXXXXXXX}

\pagerange{\pageref{firstpage}--\pageref{lastpage}} \pubyear{2015}

\maketitle

\label{firstpage}

\begin{abstract}
Classifying transients based on multi-band lightcurves is a challenging but crucial problem in the era of GAIA and LSST since the sheer volume of transients will make spectroscopic classification unfeasible. We present a nonparametric classifier that predicts the transient's class given training data. It implements two novel components: the use of the \emph{BAGIDIS} wavelet methodology -- a characterization of functional data using hierarchical wavelet coefficients -- as well as the introduction of a ranked probability classifier on the wavelet coefficients that handles both the heteroscedasticity of the data in addition to the potential non-representativity of the training set. The classifier is simple to implement while a major advantage of the \emph{BAGIDIS} wavelets is that they are translation invariant. Hence, \emph{BAGIDIS} does not need the light curves to be aligned to extract features. Further, \emph{BAGIDIS} is nonparametric so it can be used effectively in blind searches for new objects. We demonstrate the effectiveness of our classifier against the Supernova Photometric Classification Challenge to correctly classify supernova lightcurves as Type Ia or non-Ia. We train our classifier on the spectroscopically-confirmed subsample (which isn't representative) and show that it works well for supernova with observed lightcurve timespans greater than 100 days (roughly 55\% of the dataset). For such data, we obtain a Ia efficiency of 80.5\% and a purity of 82.4\%, yielding a highly competitive challenge score of 0.49. This indicates that our ``model-blind'' approach may be particularly suitable for the general classification of astronomical transients in the era of large synoptic sky surveys.
\end{abstract}


\section{Introduction}

A major shift occurring in astronomy is the dawn of wide-field synoptic surveys such as GAIA\footnote{http://sci.esa.int/gaia/} and LSST\footnote{http://www.lsst.org/lsst/} that will sweep up in their nets vast numbers of objects which then require automated sifting and classification; a trend that is pushing astronomy rapidly into the arena of machine learning.

The lure of machine learning lies in its potential to exploit large quantities of data to provide significant insights despite limited prior understanding by training on known data. This, of course, can be problematic if the training data is not representative, i.e. is systematically different in some way from the main test data that needs to be classified.  In astronomical applications, the training data is typically created by using spectra of known objects as templates and then generating ``fake'' observations of these data with noise and cadence properties appropriate for a given telescope. This is e.g. done for supernovae via the SNANA code\footnote{http://arxiv.org/abs/0908.4280}. Using simulations based on real data for training algorithms allows significant progress to be made. E.g. the same supernova spectrum can easily be simulated at a variety of redshifts, extinctions and cadence, allowing the training algorithm to be exposed to a range of examples. However, it is important to remember that such training data has fundamental limitations - it is likely to be significantly more uniform than reality since all the data come from the same underlying set of spectra and do not include the full variety of spectra likely to occur in nature. Unfortunately spectra are expensive to take for faint sources, which for supernovae in particular predominate at high redshifts, meaning that training data are unlikely to be representative but rather biased towards brighter objects and hence not representative. 

A good example of the impact of this lack of representativity is provided by the 2010 photometric supernova classification challenge \citep{Kessler2010} in which many groups competed using a number of different methods for separating the light curves of Type Ia supernovae (SNIa) from those of types Ibc and II (labelled collectively as ``non-Ia'' or ``nIa''). The training data was designed to mimic current spectroscopic followup strategies which yielded a highly biased and non-representative dataset that had a much higher Ia/nIa ratio than in the test dataset. This lack of representativity wrecked havoc with many of the classification algorithms used for the challenge. In \cite{Newling2011} it was found that a training set of 50 representative supernovae gave the same performance as a training set with approximately 1000 unrepresentative supernovae.  

While it is possible in principle to produce a representative training data, it is difficult to do so. At the faintest end even if one schedules spectroscopic followup the resulting spectrum may be too noisy to yield a clear identification. Even more tricky, one needs the training dataset to be representative for all classes of objects to be classified, which for general wide-field synoptic surveys such as GAIA, SKYMAPPER and LSST will include AGN, variables stars as well as the various sub-classes of supernovae. It is likely that a truly representative training sample for such surveys will be difficult to assemble. Therefore, it is important to develop features and classification algorithms that are relatively robust to the expected types of non-representativity in future astronomical surveys. 

In this paper we search for methods that (1) are not significantly affected by the non-representativity of the training set, (2) that do not require the light curves to be aligned, (3) do not require any redshift estimate for the object and (4) make minimal assumptions about the form of the light curves (``non-parametric'').  Photometric supernovae classification has traditionally been performed by comparing the supernova's light curve measurements to the light curve templates for the various supernova classes. The supernova class whose template most closely matched the candidate supernova's measurements would be the predicted class. However such methods are very sensitive to the error in the templates with small deviations from the true templates causing severe drops in classification performance. Recently, cosmologically-blind approaches have been proposed. These typically fit parametric models to the light curves and use the resulting parameter estimates as features that one may use to build a classifier. These approaches have been shown to work well provided the parametric form of the flux model is consistent with the flux behaviour of the transient objects being studied. As such a parametric approach -- such as that followed in \citet{Newling2011} -- is not suitable when dealing with astronomical objects whose transient behaviour is poorly understood or when the transient behaviour varies drastically between the classes of the transient object. 

Future photometric surveys will not only be detecting supernovae of course, but will also capture other transient astronomical objects -- some of whom may represent new classes altogether and which may encode much of the value of the new survey.  In such cases, parametric curves for the new classes or sub-classes do not exist and so a general nonparametric method of capturing the transient behaviour is advisable. We advocate using the coefficients of a wavelet projection as a general means of characterizing the behaviour of transient astronomical objects. See \cite{Fugal2009} for a thorough introduction to wavelets. Wavelets are a popular method of representing curves which simultaneously show localised sharp features (peaks, troughs, ...) and smooth parts. They combine the property of any nonparametric curve estimation method to take advantage of neighbouring points being more likely to have values close to each other, with the wavelet's built-in localisation: in fact the curve representation is done by adding a series of localised basis functions (e.g. of compact support). Having this property, wavelets tend to outperform principal components analysis as means of dimension reduction when applied to functional data with localised sharp transients. We thus apply a wavelet analysis to characterize the light curves, subsequently fitting various classifiers to the resulting dataset.  We also advocate the use of a ranked probability classifier as it is relatively robust to nonrepresentative training sets and can deal with heteroskedastic data. With heteroskedastic data, the astronomical measurements errors do not all share the same variance. Both these issues are relatively common in astronomical datasets.

Previous work on classifying astronomical transients from their light-curves includes the extensive work on the {\bf 500} million light curves from the Catalina Real-Time Transient Survey  \citep{Philip2012, Donalek2013, Graham2013,Djorgovski2014, Faraway2014} and the GAIA detector \citep{Long2012,Blagorodnova2014}. Wavelets have been used in regression with astronomical spectra in \cite{Machado2013} to estimate redshifts. The focus of the \citet{Machado2013} paper was to minimize catastrophic failures in the redshift estimation process. \citet{Foster1996} proposed the \emph{weighted wavelet Z-transform} as a method to identify period signals within irregularly spaced data. \citet{Graham2014} used the \emph{Slepian wavelet variance} technique to select quasars and identify their characteristic time scales.


We use wavelets to characterize functional astronomical data within the \emph{Supernova Photometric Classification Challenge} (SNPCC) dataset. The SNPCC dataset simulates flux measurements under realistic observing conditions. Consequently, the time of first detection varies greatly with some supernovae only detected well after peak-brightness. This source of heterogeneity is a problem for traditional functional data characterization as most characterization methods would consider curves that have the same shape but which are shifted with respect to each other to be very different. Indeed, dealing with shifted curves is a problem with both template fitting methods as well as the cosmology-blind parametric methods. The standard solution to the varying locations is to align the flux measurements for the various supernovae so that time zero corresponds to the estimated time at which the various supernovae reach peak brightness. This can be computationally intensive to implement and may not be appropriate for other forms of astronomical transient objects. We thus propose using the \emph{BAGIDIS} methodology to characterize the curves. \emph{BAGIDIS} is a hierarchical wavelet procedure that is robust to shifts in the positions of curves and which can also be readily applied to non-dyadic datasets -- that is datasets where the curves are of arbitrary length \citep{Timmermans2013, Timmermans2015}.

In section 2 we describe a method for extracting features from functional data. The method can be readily extended to deal with heteroskedastic data as well as functional data that is not equidistant. In section 3, we propose a classifier that is relatively robust to a non-representative training set. This is followed in section 4 by a description of the SNPCC dataset. The methods in sections 2 and 3 are applied to the SNPCC dataset in section 5 which describes the implementation of the feature extraction and classification procedures while in section 6 we compare and discuss the performances of the various implementations. Finally, we make some concluding remarks in section 7.

\section[]{Feature Extraction for Functional Astronomical Data}
Most astronomical datasets contain examples of what in the statistics literature is termed ``functional data'' where each observation consists of a set of samples of one (or perhaps several) function(s) with a single argument. For the time being, we shall assume for each of $N$ astronomical objects that there is a single corresponding function $f$ that we measure. For the $i$-th astronomical object we obtain  $n_i$ measurements of the corresponding function $f_i$ at the argument values $x_{i,1}$, $x_{i,2}$, ..., $x_{i,n_i}$ which are denoted $y_{i,1}$, $y_{i,2}$, ..., $y_{i,n_i}$. 
\begin{equation}
y_{i,j}=f_i(x_{i,j})+\epsilon_{i,j} \mbox{ \ \ where }i=1,...,N, \ j=1,...,n_i
\label{func}
\end{equation}
with $\epsilon_{i,j}$ denoting the measurement errors, which are supposed to be normally distributed $\epsilon_{i,j}\sim\mbox{Normal}(0,\sigma_{i,j}^2)$ . Examples of functional data within astronomy include astronomical spectra (here $y_{i,j}$ is the \emph{measured} light intensity of the $i$-th astronomical object at the wavelength $x_{i,j}$) as well as the light curves of variable objects such as supernovae (here $y_{i,j}$ is the \emph{measured} flux of the $i$-th supernova at time $x_{i,j}$ in a particular colour band). 

One application of functional data is in the field of classification. Suppose we have $N$ astronomical objects that are divided into $M$ classes (where $M \ll N$) and that for each object we wish to sample a function $f$ as an aid to classification. For many contexts, it is reasonable to assume that the behaviour of the function $f$ is determined by the astronomical object's class membership. Let $\tau_i$ represent the class of the $i$-th astronomical object (that is $\tau_i\in\{1,2,...,M\}$ for $i=1,...,N$). If class membership uniquely determines the functional form then eqn (\ref{func}) can be rewritten as:
\begin{equation}
y_{i,j}=f_{\tau_i}(x_{i,j})+\epsilon_{i,j} \mbox{ \ \ where }i=1, ...,N, \ j=1,...,n_i.
\label{func2}
\end{equation}
That is we have $M$ unique but unknown functions $\{f_1,...,f_M\}$ with the total number of independent samplings over these $M$ functions being $N$. Assume for each astronomical object that we have an equal number of measurements of their functions $\{f_{\tau_i}:i=1,...,N\}$. That is $n_i=n\ \ \mbox{for } i=1,...,N$. The more the functions $\{f_{\tau_i}:i=1,...,N\}$ are sampled (that is, the larger $n$ becomes), the more information there is of the behaviour of these functions for the various classes. However, if these samples are used directly as input for a classification procedure, increased sampling of the function may paradoxically decrease classification accuracy since the extra information gained from the larger $n$ may not compensate for the increased dimensionality of the input space. Increased dimensionality makes inference difficult for a number of reasons: firstly as the dimensionality of an input space increases, the expected distance between two randomly chosen sampled functions, $(y_{i,1}$, $y_{i,2}$, ..., $y_{i,n})$ and $(y_{j,1}$, $y_{j,2}$, ..., $y_{j,n})$ increases. That is, data becomes more sparse as the dimensionality of the input space increases. Furthermore the complexity of the classification rule often increases with the dimensionality of the input space thus dramatically increasing the variance of the estimated classification boundary.

Consequently, it is worthwhile to seek to reduce the dimensionality of the input space whilst preserving as much information within the dataset as possible. The traditional and most popular method for dimension reduction is principal components analysis (PCA). Here, the $N$ observations of the $n$ correlated samples are transformed into a set of (at most) $n$ linearly uncorrelated principal components. Principal components analysis (PCA) has been used to successfully compress stellar spectra by a factor of over 30 \citep{Bailer1998}. The reduced dataset was then fed into a neural network for classification. Instead of using PCA for dimension reduction, we instead propose using wavelets for performing dimension reduction. Wavelets are uniquely tailored to represent functional data whose underlying signal consists of a time-varying mixture of frequency components. Such functional datasets are prevalent in the astronomical literature and hence wavelets are a natural tool for analysis of many astronomical datasets. In the next subsection we propose a method of wavelet analysis that can be readily applied to a wide range of functional astronomical data.


\subsection{The \emph{BAGIDIS} Methodology}

In our initial discussion of the \emph{BAGIDIS} methodology we will make two main assumptions that will subsequently be relaxed in forthcoming sections. Firstly, we assume for every astronomical object that there is a corresponding function $f_i$ that we wish to estimate and that all $N$ of these functions are sampled on the same equally-spaced grid i.e. $n_i=n=n_j$ with $(x_{i,1}$, $x_{i,2}$, ..., $x_{i,n})=(x_{j,1}$, $x_{j,2}$, ..., $x_{j,n}) \mbox{ for } i,j\in\{1,...,N\}$ and also that $x_{i,(k+1)}-x_{i,k}=\Delta=x_{j,(l+1)}-x_{j,l} \mbox{ where } i,j\in\{1,...,N\} \mbox{ and } k,l\in\{1,...,n-1\}$. Secondly, we assume that the measurements error variances are the same i.e. $\sigma_{i,j}^2=\sigma^2=\sigma_{i^*,j^*}^2$ for $i,i^*\in\{1,...,N\}$ and $j,j^*\in\{1,...,n\}$. As pointed out in the introduction to this section, we would like to benefit from the ability of wavelet methods to compress the significant information of curves into a small number of relevant features \citep{Antoniadis2009}, used for subsequent (functional) classification. 

The higher the compression (i.e. the reduction of the dimensionality) -- without losing significant information -- the better for the subsequent classifier. However, in this context and in fact more generally, a recurring problem in functional data analysis is to estimate the similarity of various curves. Traditional methods of estimating curve similarity tend to be very sensitive to shifts in the observed curves. As an example, if two curves have exactly the same shape, but one of the curves is shifted by an amount $H$ from the other, then many measures of curve similarity will consider the two aforementioned curves to be very dissimilar. This problem is particularly relevant to astronomical data sets, as e.g. astronomical spectra as well as the SNPCC dataset, to be analysed subsequently here: the onset of a peak in the spectrum or in a light curve can be slightly shifted from one data set to another without making the two curves necessarily belong to different classes.


In the case of supernova classification, previous approaches have circumvented the aforementioned problem by first aligning the sampled curves. In the case of supernova classification, this alignment is often achieved by first attempting to find the time at which each light curve achieves peak brightness and subsequently shifting the time axis so that time zero corresponds to the time of peak brightness \citep{Newling2011,Ishida2013}. Though such an alignment procedure works well in practice, there are a number of disadvantages to such an approach. Firstly, implementing such a procedure is considerably more difficult if the function has several peaks. This is not the case for the various supernova classes, however one can expect future photometric surveys to find transient objects that have multiple peaks and so aligning the curves may be difficult. Secondly, the alignment procedure could be computationally intensive. Thirdly, many alignment procedures make implicit assumptions about the form of the functional data. Such assumptions may not necessarily be valid for all classes of astronomical objects. We thus choose to implement a comparatively new wavelet procedure -- called \emph{BAGIDIS} \citep{Timmermans2013} -- that explicitly accounts for possible shifts in the curves. Furthermore, unlike traditional wavelet analysis, \emph{BAGIDIS} does not require the dataset to be dyadic. This is important as the sequence of measurements for the curve will often not be dyadic. Finally, as a key to success, \emph{BAGIDIS} uses a hierarchical ordering of its wavelet coefficients in order to encode ``optimally'' the significant features of various, potentially misaligned curves with the purpose of discriminating them by a small number of coefficients. We describe the \emph{BAGIDIS} methodology in more detail now.

The acronym \emph{BAGIDIS} stands for BAses GIving DIStances, as basis expansion is at the core of
the methodology, together with the introduction of a new distance measure in the coefficient domain of this expansion. The \emph{BAGIDIS} methodology has been introduced in \citet{Timmermans2015}, where significant gains in performing dimension reduction over other methods, such as PCA, have been demonstrated (e.g. in a blood-serum analysis of spectra). Further investigation
of its use in nonparametric functional prediction and classification \citep{Ferraty2006} is provided
in \citet{Timmermans2013}. The following description of the key ideas is general enough for application of the method to any set of curves with one or more localised features (spikes, peaks, troughs, discontinuities,...) such as light curves and spectra, not just those resembling the supernova light curve dataset that we shall subsequently use to demonstrate the methodology which are of relatively short length and which show at most two localised humps. 

As a first step of the procedure, if necessary by the length of the data, each curve is decomposed in a set of short series using a sliding window, so that each windowed series should not contain two significant patterns of the same amplitude (in our specific application this was not necessary). Subsequently the (windowed) series is expanded in a particular wavelet basis, with wavelet coefficients called $d_k$, which are referred to as the ``Unbalanced Haar Wavelet Basis'' (UH-basis\footnote{In the literature, the UH-basis is called the ``Best Suited Unbalanced Haar Wavelet Basis'' (BSUHWB). We instead use ``UH-basis'' for simplicity.}). It is obtained using the ``Unbalanced Haar Wavelet Transform'' (UH-WT\footnote{Similarly, the UH-WT is referred to as the ``Bottom-Up Unbalanced Haar Wavelet Transform'' (BUUHWT).}) proposed by \citet{Fryzlewicz2007}. This general version of \emph{BAGIDIS} -- where each curve is decomposed into a set of short series -- offers the possibility to localise the analysis along shorter windows. This decomposition prevents otherwise similar features that are not just slightly shifted but are far apart from each other from being treated as the same feature (in case this is undesirable from the nature of the problem at hand). 

\begin{figure*}
\includegraphics[width=17.8cm]{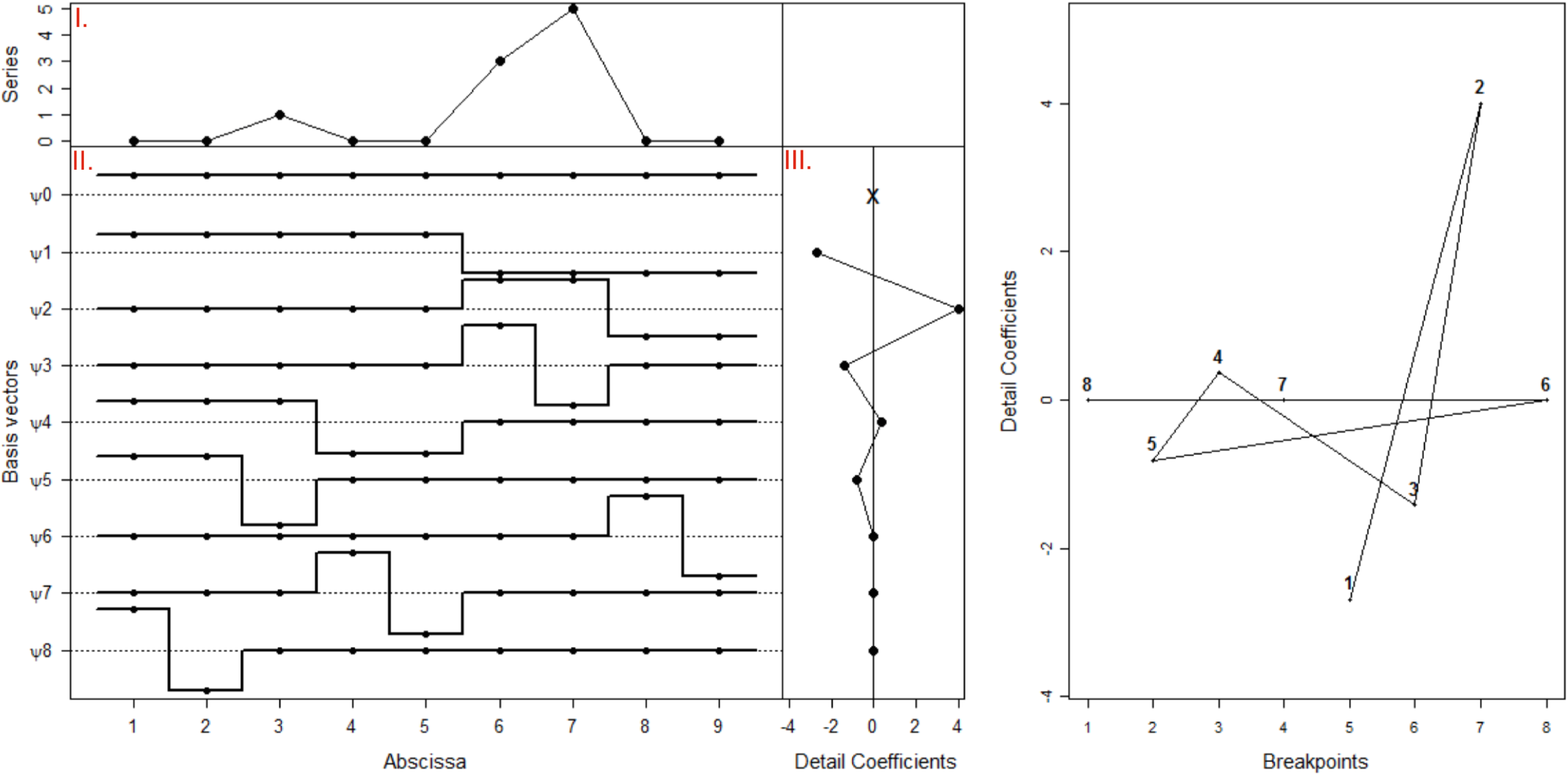}
\caption{Left panel: Illustration of a UH-WT expansion. The left panel consists of three parts denoted I., II., and III. respectively. In the upper part of the left panel (part I.) we plot the series. The corresponding abscissa axis at the very bottom is common for that graph and for the graph of basis vectors (part II.). The main part of the figure (part II.) shows the basis vectors of the Unbalanced Haar wavelet basis that is best suited to the series (UH-WT basis). These vectors are represented rank by rank, as a function of an index along the abscissa axis and are denoted $\psi_0$, ..., $\psi_8$ where the subscript denotes the rank. Dotted horizontal lines indicate the level zero for each rank. Vertically, from top to bottom on the right hand side (part III.), we find the detail coefficients associated with the wavelet expansion. Each coefficient is located next to the corresponding basis vector. Right panel: Representation of the series in the breakpoint-wavelet coefficient (b-d) plane. The same series is plotted in the plane that is defined by the values of its breakpoints and its detail coefficients. Points are numbered according to their rank in the hierarchy.}
\label{Bagidis} 
\end{figure*}

Haar wavelets are made of level-change indicator functions on a hierarchy of dyadic subintervals \footnote{These are partitionings of an interval into halves, quarters, eights etc.} of the unit (observation) domain, they are piecewise constant to the left and the right of the level-change point and normalised such that they constitute an $L_2-$ orthonormal basis (akin to the Fourier basis but now being localised on short intervals and hence comprising much better compression properties).  {\em Unbalanced} Haar wavelets overcome the dyadic restriction: the level-change (or breakpoint) $b_k$ can be at any of the observation points of the underlying signal. Still they constitute an orthonormal basis when implemented via the UH-WT: in fact, the set $\{(b_k,d_k)\}_k$ determines the shape of the projected signal uniquely, and induces an interesting property of hierarchy in the resulting UH-basis expansion. The most important level change of the signal, i.e. its location and its amplitude, is encoded in $(b_1,d_1)$, whilst secondary local features are to be found in higher coefficients.  An illustration of this along a simple noiseless example made of 9 data points is shown in Fig. \ref{Bagidis}. As hoped for and observed by looking at the location of the discontinuity points $b$ between positive and negative parts of the wavelets, the first (three) non-constant vectors support the largest level-changes of the series and encode therefore the highest peak of the series. The subsequent (two) vectors point to smaller level changes (encoding the small peak) while the last few vectors correspond to zones where there is no level change -- as indicated by the associated zero coefficients.

In a noisy set-up, the UH-WT coefficients with ``ranks'' (i.e. indices) $k$ higher than some integer $D$ do not carry any more significant information (but just noise contribution) and should be suppressed from the basis expansion. (The critical rank $D$ is usually small but needs to be user-chosen -- ideally in a data-driven way e.g. cross-validation -- as depending on the problem at hand.)
In such a way, the UH-WT allows for an automatic and unique hierarchical description of each of the curves into a curve-adapted orthonormal basis. The hierarchy makes the resulting bases (UH-basis) comparable to each other, although different, and allows subsequent classification based on a distance of similarity of two measured curves $\mathbf{y}^{(1)}$ and $\mathbf{y}^{(2)}$ -- each of which have been sampled $n$ times -- via similarity of their $\{(b_k,d_k)\}_{k=1}^D$ coefficients (where $D\leq n$):
\begin{equation}
d(\mathbf{y}^{(1)}, \mathbf{y}^{(2)})= \sum_{k=1}^D w_k\ \left(\alpha |b_k^{(1)}- b_k^{(2)}|^2 + (1-\alpha) |d_k^{(1)}- d_k^{(2)}|^2  \right)^{1/2}
\label{dist}
\end{equation}
Here the tuning parameter $\alpha \in [0,1]$ can be chosen according to the application at hand (or even by supervised learning). In cases where the astronomical curves are shifted with respect to each other yet where we choose to avoid any aligning of the curves, the \emph{BAGIDIS} procedure is still able to recognize that the curves are the same since the hierarchical ordering of the ranks is performed in a data-driven manner. Hence, by putting all the weight in eqn \ref{dist} on the $d_k$'s (i.e. $\alpha=0$), two curves that are shifted with respect to each other will be considered identical - this is often desirable for classification. Indeed, for the rest of the paper we take $\alpha=0$ as a principled choice.  With the weight function $w_k$ one has the flexibility to give more weight on more important features (which are encoded in the first coefficients). Again this parameter can be chosen in a supervised way, e.g. by cross-validation.

Selecting the optimum number $D$ of \emph{BAGIDIS} coefficients for classification involves a tradeoff: the more coefficients we use to represent the dataset, the smaller the information loss. However, more coefficients will cause the classifier to have higher variance. The optimum number of \emph{BAGIDIS} coefficients to train a classifier can be selected from the data, by e.g. a  cross-validation procedure, by choosing the number of \emph{BAGIDIS} coefficients that minimize the misclassification error rate for the validation sets. A \emph{BAGIDIS} package has been developed for \emph{R} and is available on the CRAN website \footnote{http://cran.r-project.org/web/packages/Bagidis/}. As the package stands, the order of the algorithm is $O(n^2)$ in the number of light curve points. In future large synoptic sky surveys, $N$ will increase by at least two orders of magnitude, but $n$ will only increase moderately due to more regular sampling and longer monitoring. If $n$ proves to be too large, one may decompose a curve into a set of short series using a sliding window. Stochastically varying objects, such as AGN, will not have light curves well-described by the lowest few wavelet coefficients, and it is unclear how well they will be classified using our methodology. This is a topic left for future work but is not important in the context of supernovae since AGN and other long-term variable objects are typically excluded up front by virtue of their history of variability.


\subsection{Estimating the Variances of the BAGIDIS Coefficients}
In the previous section a distance measure (given by eqn \ref{dist}) was proposed. Note that since \emph{BAGIDIS} imposes a hierarchical ordering upon its wavelet coefficients, we'd typically expect the weights $w_k$'s to be a monotonically decreasing function of $k$. \citet{Timmermans2015} proposed that the weights be given by:
\[
w_k=\frac{\log(n+1-k)}{\log(n+1)}.
\]
where $n$ is the number of measurements of the function $f$. It was found that this functional form worked well for many datasets. However whatever the functional form chosen, the weights (as shown in eqn \ref{dist}) do not depend on the pair of observed curves $(\mathbf{y}^{(1)}, \mathbf{y}^{(2)})$ for which we wish to calculate a distance measure. This simplified structure is natural if the input is homoskedasitc. Indeed, homoskedasticity is a standard assumption for most traditional classifiers. However for astronomical data, there will often be a large level of heterogeneity in the variances of the wavelet coefficients. This heterogeneity stems from the underlying heterogeneity of the measurement process: both the preciseness and the number of measurements collected for the pool of astronomical objects within a database can commonly vary by at least an order of magnitude. This heterogeneity must be accounted for within the distance measure. 

Suppose that the variances of the $k$-th \emph{BAGIDIS} coefficients for the observations $\mathbf{y}^{(1)}$ and $\mathbf{y}^{(2)}$ are denoted by $\sigma_{k}^{(1)^2}$ and $\sigma_{k}^{(2)^2}$ respectively. If these variances are known, then a more appropriate distance measure would be:
\begin{multline}
d(\mathbf{y}^{(1)}, \mathbf{y}^{(2)})= \sum_{k=1}^D \bigg[\frac{w_k}{\sqrt{\sigma_{k}^{(1)^2}+\sigma_{k}^{(2)^2}}}\ \Big(\alpha |b_k^{(1)}- b_k^{(2)}|^2 \\
+ (1-\alpha) |d_k^{(1)}- d_k^{(2)}|^2  \Big)^{1/2}\bigg]
\label{dist2}
\end{multline}
We thus turn our attention to the estimation of the variances $\sigma_{k}^{(1)^2}$ and $\sigma_{k}^{(2)^2}$ of the \emph{BAGIDIS} coefficients.


For astronomical datasets, each measurement on an object often includes a corresponding estimate of the variance of its measurement error. In principal it could be possible to derive an analytical expression for how the variance of the measurement errors filters through to a set of wavelet coefficients. However, such an approach is not feasible for the \emph{BAGIDIS} coefficients as the underlying wavelet coefficients are re-ordered in order of importance. This re-ordering makes the derivation of an analytical formula for the variances of the \emph{BAGIDIS} coefficients very difficult. 

Instead, we wish to estimate the variances by Monte Carlo simulation. Assume the measurement errors $\epsilon_{i,j}$ for the functions $\{f_i:i=1,...,N\}$ are Gaussian with mean zero and known (possibly unequal)  variances $\sigma^2_{i,j}$. Under this assumption we may use Monte Carlo simulations to estimate the variances of the \emph{BAGIDIS} features. This entails using the distributional information on the measurement errors to resample a new set of equidistant measurements for each function. By subsequently applying the \emph{BAGIDIS} procedure to each simulated set of equidistant measurements, we can investigate how the randomness of the measurement errors feeds through to the \emph{BAGIDIS} coefficients. By simulating several sets of \emph{BAGIDIS} coefficients for each astronomical object, we create a pool of coefficients that we can use to precisely estimate the variances of the \emph{BAGIDIS} coefficients for each astronomical object. This variance estimation procedure will be modified in section 2.3 to allow for non-equidistant measurements. In cases where the resulting simulated \emph{BAGIDIS} coefficients may have outliers, one can use M-estimators instead of the sample variance as an estimator of the true variance. See Chapter 5 of \citet{vanderVaart2000} for more details about the M-estimators.

\subsection{Allowing for Non-Equidistant Functional Data}
Given a sequence of $n$ equidistant observations, the \emph{BAGIDIS} methodology will return $n$ wavelet coefficients that are ordered according to the importance of the sequence's level changes. Thus in order to use \emph{BAGIDIS} to compare the similarity between pairs of astronomical objects, we require their corresponding functions to each be sampled on the same equidistant grid. We assumed the existence of such a grid in sections 2.1 and 2.2. However, this assumption is rarely met by astronomical datasets since observing time on instruments are limited and instruments are often subject to adverse observing conditions. Consequently, it is important that we are able to adapt any arbitrary sampling of the function of interest into a format that is usable by the \emph{BAGIDIS} methodology. We propose the utilization of splines to create the required dataset. Fitting a natural cubic spline is less computationally expensive than Gaussian process regression (an alternate method to perform nonparametric regression). If instead of conducting a functional analysis of a signal, one were interested in characterizing periodic or quasi-periodic signals, one may use alternative wavelet transforms that do not require the data to be regularly spaced \citep{Foster1996,Graham2014}. In our case however, we require a functional analysis of the underlying signal via encoding of its local level shifts as an aid to classification.

Suppose we have $N$ astronomical objects sampled at the points $\{(x_{i,1}$, $x_{i,2}$, ..., $x_{i,n_i}): i=1,...,N\}$ and suppose that $x_{i,1}=0$ and $x_{i,n_i}\geq T$ for all $i$. That is, for all $N$ astronomical objects we have measurements that span the interval $[0,T]$. Under such conditions, it is possible to fit a spline to each object over the region $[0,T]$. The spline estimates at points along an equidistant grid within $[0,T]$ can subsequently be used as the input for the \emph{BAGIDIS} procedure. 

Given a sequence of measurements $\{y_{i,j}:j=1,2,...,n_i\}$ at times $\{x_{i,j}:j=1,2,...,n_i\}$ for the $i$-th object, we wish to find a function $g_i(x)$ that will fit this data well (that is have a small residual sum of squares) whilst also being reasonably smooth. A natural cubic spline -- with knots at the points  $(x_{i,1}$, $x_{i,2}$, ..., $x_{i,n_i})$ -- will be the function $g_i(x)$ that minimizes the following objective:
\begin{equation}
\sum_{j=1}^{n_i}{(y_{i,j}-g_i(x_{i,j}))^2}+\lambda_i\int_x{g_i''(x)^2dx}.
\end{equation}
The first term is the sum of squared residuals for the cubic spline whilst the second term is a regularization factor that penalizes roughness within $g_i(x)$. $\lambda_i$ represents a tuning parameter for the $i$-th object that controls the trade-off between the fidelity to the data and the roughness of the spline estimate. Since astronomical data is typically heteroskedastic (that is the measurement errors do not all share the same variance), we instead find the spline that minimizes:
\begin{equation}
\sum_{j=1}^{n_i}{\frac{1}{\sigma_{i,j}^2}(y_{i,j}-g_i(x_{i,j}))^2}+\lambda_i\int_x{g_i''(x)^2dx}
\label{weightSpline}
\end{equation}
where as before $\sigma_{i,j}^2$ is the variance of the $j$-th measurement error for the $i$-th object. As such, the resulting splines tends to follow the precisely measured fluxes more closely, thus making the method relatively robust to outliers resulting from imprecise measurements.

We choose the value of the tuning parameter $\lambda_i$ for each astronomical object (and hence our cubic spline function $g_i^{\lambda_i}$) by cross-validation. \emph{Leave-one-out} cross-validation is calculated by:
\begin{equation}
CV(g_i^{\lambda_i})=\frac{1}{n_i}\sum_{j=1}^{n_i}{\left(\frac{y_{i,j}-g_i^{\lambda_i}(x_{i,j})}{\sigma_{i,j}(1-\mathbf{S}_{\lambda_i}^{(i)}(j,j))}\right)^2}
\end{equation}
where $\mathbf{S}_{\lambda_i}^{(i)}$ is the \emph{smoother matrix} (for details on how to calculate the \emph{smoother matrix} refer to sections 5.2.1 and 5.4 of \citet{Hastie2009}). $\lambda_i$ is chosen so as to minimise $CV(g_i^{\lambda_i})$.

If we wish each astronomical object to be sampled on an equidistant grid of length $n^*$, we can set $x_{i,1}^*=0$, $x_{i,2}^*=\Delta$, $x_{i,3}^*=2\Delta$, ..., $x_{i,n^*}^*=(n^*-1)\Delta$ for $i=1,2,...,N$ with the sampled values given by the corresponding spline estimates i.e. $y_{i,1}^*=g_i(0)$, $y_{i,2}^*=g_i(\Delta)$, $y_{i,3}^*=g_i(2\Delta)$, ..., $y_{i,n^*}^*=g((n^*-1)\Delta)$. This sequence of $n^*$ spline estimates can be taken as input for the \emph{BAGIDIS} procedure. The `$^*$' on each variable denotes that the variable is part of a data sequence that has been adapted to be equidistant. $\Delta$ is a tuning parameter whose value we choose. When an underlying function $g(x)$ is sampled at intervals of length $\Delta$, there are an infinite number of continuous functions that will fit the sampled points. However, only one of them will not have any frequency components higher than $1/(2\Delta) \mbox{ days}^{-1}$. Since the \emph{BAGIDIS} procedure approximately reconstructs this bandlimited function, $\Delta$ must be set low enough so that we don't lose any important high frequency components of the underlying signal. On the other hand the smaller $\Delta$ is, the more computationally expensive will be the feature extraction.

In order to get the variances of the \emph{BAGIDIS} coefficients, we again use Monte Carlo simulation: for each astronomical object, we simulate a set of measurements. For each set of simulated measurements, we can fit a spline using the inverse variance weighting, sample the fits on an equidistant grid and use the resulting sequence as input for the \emph{BAGIDIS} procedure. Fig. \ref{splines1000} shows the resampled splines for the flux measurements of a supernova with the accompanying description and discussion given in section 5.1. The estimated variances of the \emph{BAGIDIS} coefficients can consequently be used to calculate a dissimilarity measure as given by eqn \ref{dist2}. By using this suggested procedure, the \emph{BAGIDIS} methodology can be used to generate a $N \times D$ features set for many types of functional data -- even those for which the sampling is unequally spaced and the measurements are heteroskedastic. Typically $D \ll n_i$ for $i=1,...,N$ where as before $n_i$ is number of equidistant observations for the $i$-th observation and so the feature set will be in a far lower dimensional space than the original input whilst also minimizing information loss.

\section[]{A Ranked Probability Classifier}
Suppose we have $N$ astronomical objects whose class $\tau$ is known (the training set) and an additional $L$ objects of unknown class (the test set). Ideally the properties of the training set should match those of the test set. However for astronomical datasets, the training set will often have a different distribution (e.g. lower redshift, lower apparent magnitude) than the test set. This difference in distribution is problematic as many classifiers are sensitive to non-representativity. One approach to making a classifier more robust to non-representativity is to make pairwise comparisons between the candidate from the test set and each object in the training set. The predicted class for the candidate will then only be a function of the $K$ closest matches within the training set where $K$ can be chosen by cross-validation. This subset of the training set are the only objects that influence the predicted class and they will typically be more representative of the test set when the input space is low-dimensional. The most common example is the $K$ nearest neighbour classifier (kNN). \citet{Ishida2013} successfully applied kNN to classify supernovae light curves with a highly non-representative training set. However, kNN cannot give the class probabilities and does not naturally account for the heteroskedacity of the dataset. We thus develop a ranked classifier that can address these two shortcomings.

Assume for all $N+L$ objects that we have obtained a sequence of $n$ measurements of a function $f$ (as in eqn \ref{func2}) on an equidistant grid. Hence the information on the function $f$ -- compressed into a vector $\mathbf{d}=(d_1,d_2,...,d_D)$ of \emph{BAGIDIS} coefficients -- can be used to classify the $L$ test objects of unknown type. The proposed ranked probability classifier works very similarly to kNN. For each candidate object within the test set, we wish to calculate the probability that its observed vector $\mathbf{d}_{\mathcal{C}}$ could have arisen from measurement error if the candidate in reality shared the same underlying function as (say) the $i$-th object within the training set. Assuming Gaussianity and independence, the relevant density is given by:
\begin{multline}
g(\mathbf{d}_{\mathcal{C}}|\mathbf{d}_i)=(2\pi)^{-(D/2)}|\Sigma_{\mathcal{C}}+\Sigma_i|^{-1/2}\\
\exp\left\{-\frac{1}{2}(\mathbf{d}_{\mathcal{C}}-\mathbf{d}_i)^T(\Sigma_{\mathcal{C}}+\Sigma_i)^{-1}(\mathbf{d}_{\mathcal{C}}-\mathbf{d}_i)\right\}
\end{multline}
where $\Sigma_{\mathcal{C}}=\mbox{diag}\big({\sigma_1^{(\mathcal{C})}}^2,...,{\sigma_D^{(\mathcal{C})}}^2\big)$ and $\Sigma_i=\mbox{diag}\big({\sigma_1^{(i)}}^2,...,{\sigma_D^{(i)}}^2\big)$. We wish to calculate this density for all objects in the training set i.e. $\{g(\mathbf{d}_{\mathcal{C}}|\mathbf{d}_i):i=1,...,N\}$. The candidate is then classified as being of the same class as the training set object that yields the highest density value. One can also calculate the approximate probability of the candidate object belonging to a class $m$ using:
\[
\mbox{Pr}(\tau_{\mathcal{C}}=m)\approx\frac{g(\mathbf{d}_{\mathcal{C}}|\mathbf{d}_m^*)}{\sum_{i=1}^{M}{g(\mathbf{d}_{\mathcal{C}}|\mathbf{d}_i^*)}}, \ \ \  m=1,...,M
\]
where $\mathbf{d}_i^*$ is the \emph{BAGIDIS} vector of the closest matching object within the training set that belongs to class $i$. Though we are able to calculate the class probabilities, this ability was only possible through assuming the coefficients came from an independent Gaussian distribution. \emph{BAGIDIS} coefficients often have a Gaussian distribution or can easily be transformed to a Gaussian distribution. Unfortunately the independence assumption is highly unlikely to be met due to the re-ordering of the coefficients. That said, the classifier seems to be relatively robust to the violation of the independent Gaussian assumption.

In order to classify each of the $L$ candidate objects, a brute-force approach would calculate the dissimilarity between a candidate object and each of the $N$ training objects. Hence the classification of the $L$ objects will require $O(L\times N)$ dissimilarity calculations. Though with moderately sized datasets -- such as the Supernova Photometric Classification Challenge dataset (SNPCC) in section 4 -- a brute-force implementation is straightforward, upcoming large synoptic sky surveys will make such an approach more difficult. We note however that it is trivial to construct a semi-metric from the Gaussian density. This semi-metric does not satisfy the triangle inequality. Nonetheless, efficient approaches to semi-metric dissimilarity searches have been developed, such as \cite{skopal2006fast}. This technique can be summarised as follows, first a transformation is applied to the semi-metric such that the triangle inequality is approximately enforced. The degree to which it is enforced is traded off against the intrinsic dimensionality of the data set (a measure of how efficiently the data set can be searched in principle). Next, one of a number of Metric Access Methods (MAMs) can be applied to search for the nearest object to the candidate object. Examples include algorithms such as the Linear Approximating Eliminating Search Algorithm \citep{chavez2001searching}. The efficiency of the search using a MAM should be far better than the brute force approach with only a small trade off in terms of accuracy of finding the closest matching object. A detailed exploration of computationally efficient implementations of the ranked probability classifier utilising these techniques is an area for future research.


\section[]{The Supernova Photometric Classification Challenge (SNPCC) Dataset}
For the purpose of this paper we use the \emph{Supernova Photometric Classification Challenge} (SNPCC) dataset\footnote{The data is available at: http://www.hep.anl.gov/SNchallenge/}. The SNPCC dataset consists of simulated light curves for over 18,000 supernovae. A subset of these simulated supernovae are considered to be spectroscopically observed \citep{Kessler2010}. The spectroscopically confirmed subset were based on which of the simulated light curves could be observed by either a 4 m class telescope with a limiting R-band magnitude of 21.5 or an 8 m class telescope with a limiting i-band magnitude of 23.5. Less than 7\% of the simulated supernovae would be observable by either of the two telescopes and hence can be considered to be spectroscopically observed. This subset are the supernovae for which the supernova class is known. The challenge was released in January 2010 and consisted of trying to develop a classifier based on the spectroscopic subset (the training set) that would accurately predict the class membership of the rest of the dataset (the test set) using the supernova's light curve data. In particular, the focus is on correctly identifying the type Ia supernova within the test dataset.

\begin{figure}
\includegraphics[width=8.5cm]{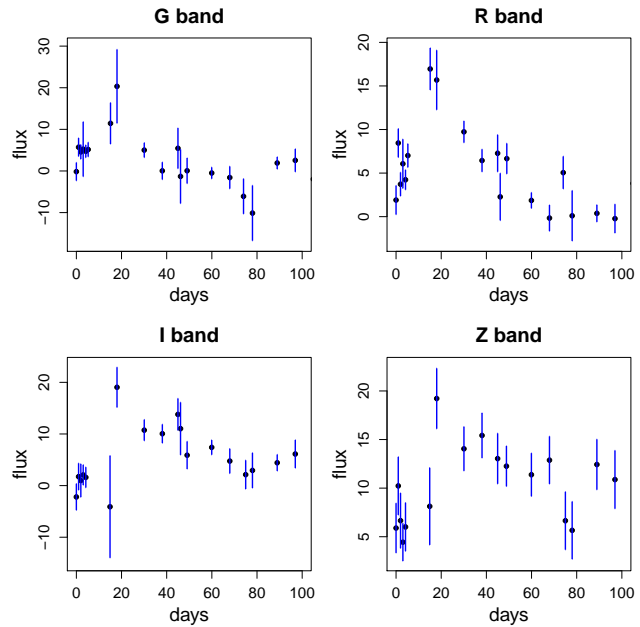}
\caption{Flux measurements with corresponding error bars over four colour bands for a randomly selected supernova within the SNPCC dataset.}
\end{figure}

For each supernova, we have a series of flux measurements over four frequency bands (G-, R-, I- and Z-bands) together with both the time of the measurement as well as the corresponding flux measurement errors. Fig. 2 shows the measurements recorded for a randomly selected supernova. These light curve measurements were simulated under a variety of observing conditions. Consequently the datasets are highly variable: the number of measurements taken and the flux measurement error vary greatly from one supernova to the next. Fig. 3 shows the distributions of  both the number of flux measurements as well as the observed timespans (elapsed time between the final and the first flux measurement of the supernova) in the G-band for the combined SNPCC dataset. The top panel is clearly bimodal with a high proportion ($\sim40\%$) of the supernovae having less than 20 flux measurements taken. The number of flux measurements is strongly correlated with the supernova's time span ($\rho\sim0.9$) with roughly the same proportion of supernovae having an observation time span of less than 100 days. In contrast, future surveys such as LSST will effectively monitor all supernovae with high cadence over their entire lifespan from before peak brightness to well after 100 days. In the SNPCC dataset, the times at which measurements are taken are irregular and differ between supernovae. These highly variable datasets are not suitable for many traditional methods of analysis that often assume that the data is standardized. It is thus important to find a general method -- such as that proposed within section 2 -- to extract a standardized set of features for each of the supernova.

\begin{figure}
\includegraphics[width=8.5cm]{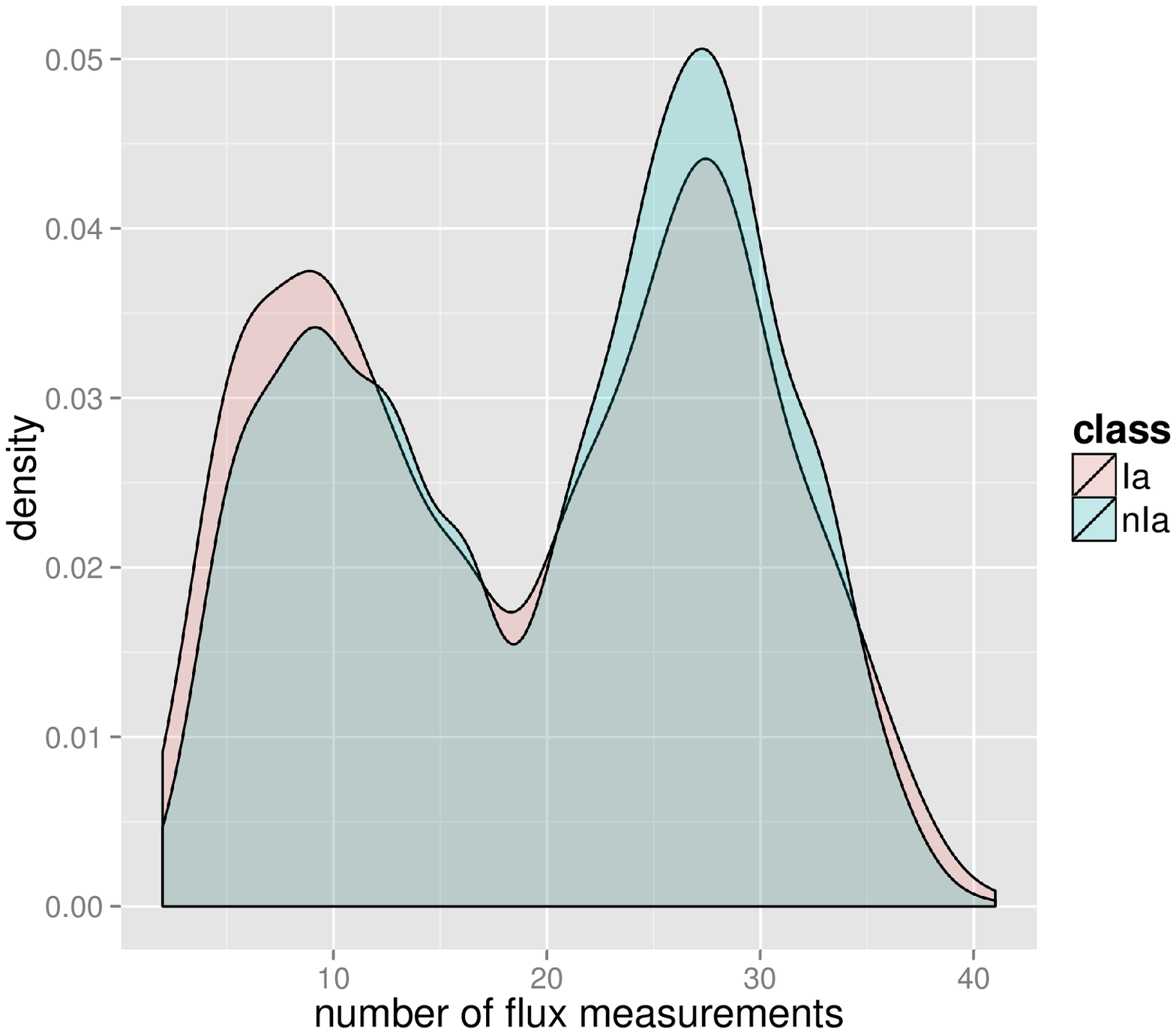}
\includegraphics[width=8.5cm]{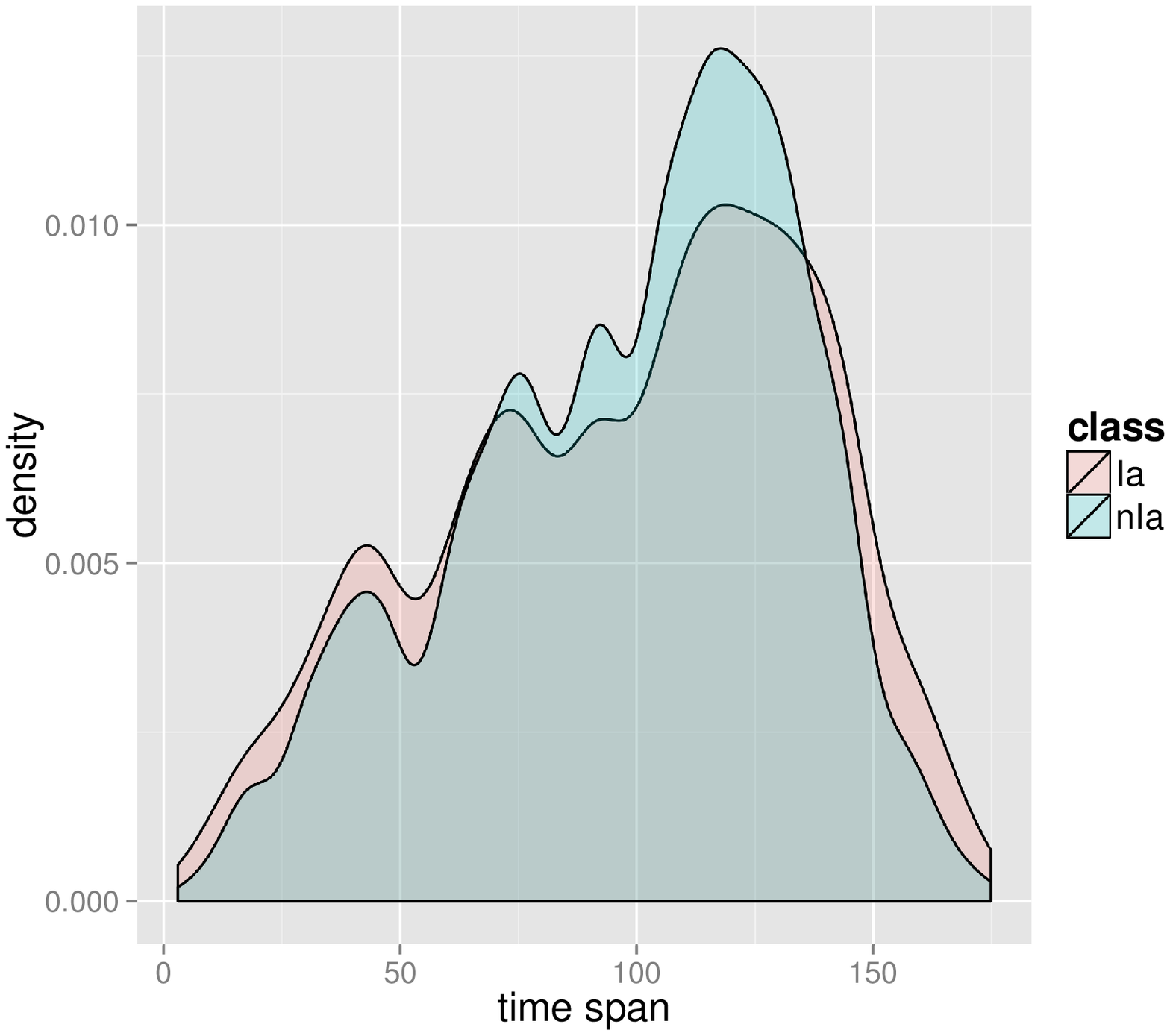}
\caption{The top panel is a density plot of the number of flux measurements in the G-band taken for each supernova within the aggregated dataset (i.e. both training and test sets). The bottom panel shows a density plot of the number of days that have elapsed between the first and the last observation for each supernova. The red plots are for the type Ia supernovae and the blue plots are for the non-Ia's. The measurement process appears to be roughly the same for both classes of supernova.}
\end{figure}

With the SNPCC dataset, the training set is also highly unrepresentative of the test set. The non-representativity arises due to the procedure for spectroscopic follow-up. Since spectroscopic followup is expensive, the training set tends to consist of supernova with much higher apparent magnitudes (see Fig. 4). Also since type Ia supernova are generally brighter than non-Ia's, the proportion of Ia's within the training set ($\sim 70\%$) is much higher than the proportion of Ia supernovae within the test set ($\sim 30\%$). In the case of the non-Ia supernovae, the test set will tend to have a pool of supernovae that are either intrinsically dimmer or that are of higher redshift. We thus need to develop a classifier that is not overly sensitive to the non-representativity of the training set. These two aforementioned problems of non-standardized datasets and non-representativity shall be addressed using the methods proposed within sections 2 and 3. We discuss the application of these methods to the SNPCC dataset in the next section.

\begin{figure}
\includegraphics[width=8.5cm]{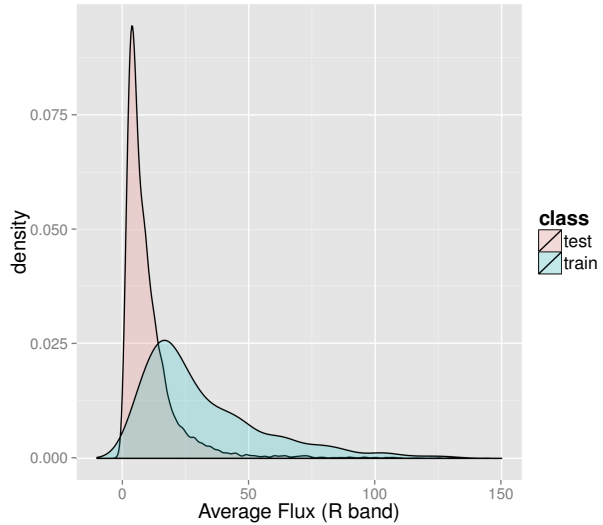}
\caption{Density plot of the average measured flux in the R-band for each supernova. The red density plot is for the test set and the blue plot is for the training set. The supernovae within the training set tend to be considerably brighter than the supernovae within the test set.}
\end{figure}

\section[]{Application of the methodology to the SNPCC dataset}
\subsection{Feature Extraction for the SNPCC dataset}
\begin{figure}
\includegraphics[width=8.5cm]{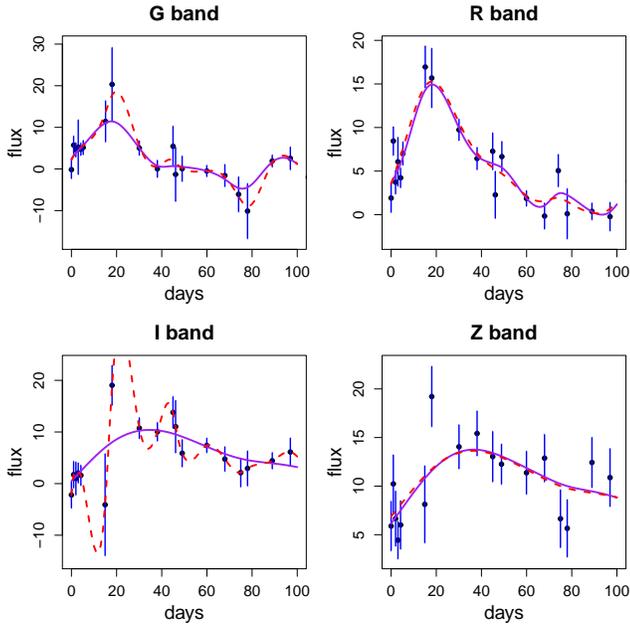}
\caption{Observed flux measurements for a randomly chosen Ia supernova over four frequency bands. In all four panels, two cubic splines are fitted to the observations. The dashed lines is a standard cubic spline whilst the solid line represents the cubic spline that results when the flux measurements are weighted inversely by the variance of the measurement error. The standard spline fit is particularly unstable for the I-band as it does not account for the large error bars associated with the outliers. The weighted spline takes the measurement errors into account and thus provides the more stable fit.}
\label{splinecomp}
\end{figure}

In order to deal with the wide variation in both the number of flux measurements as well as the observed time spans for the supernovae within the SNPCC dataset, we wish to obtain for each supernova a set of flux estimates on an equidistant grid. For the SNPCC analysis, we require flux estimates at the grid values $(0 \mbox{ days}, 2 \mbox{ days}, 4 \mbox{ days}, ..., 98 \mbox{ days}, 100 \mbox{ days})$. 
We chose a 2 day sampling frequency (i.e. $\Delta=2$) as we don't expect there to be any frequency components higher than $1/(2\Delta)=1/4 \mbox{ days}^{-1}$. Also, having $\Delta$ as small as 2 still remains computationally feasible. We use cubic splines (section 2.3) to estimate the flux values at the required times, with the weights of each measurement being inversely proportional to its variance (see eqn \ref{weightSpline}). Fig. \ref{splinecomp} demonstrates how the inverse weighting results in more stable spline fits. The regularisation parameter is chosen by cross-validation, with the parameter constrained to lie on the interval $0.45\leq\lambda\leq0.9$. This restriction assumes that the underlying function that we are sampling is slowly varying (due to the lower bound) and that it is probably nonlinear (due to the upper bound). For the majority of the supernovae, the cross-validated value of $\lambda$ lies well within the interval. For the remaining supernovae, the cross-validated value of $\lambda$ will either be equal to 0.45 or equal to 0.9 and the spline fit will -- like the majority of supernovae with $\lambda$ between 0.45 and 0.9 -- have the desired rapid rise to peak brightness before gradually decaying.

For each supernova, we have four sequences of 51 measurements: these are the spline estimates at $(0 \mbox{ days}, 2 \mbox{ days}, ..., 100 \mbox{ days})$. We wish to compress the resulting 204-dimensional space into a smaller subspace so as to reduce classification error. For each sequence, we compress the sequence into $D$ coefficients using the \emph{BAGIDIS} methodology. The coefficients are hierarchical with the first \emph{BAGIDIS} coefficient in each sequence storing the sequence mean. Since the training data is brighter than the test data, we ignore the first coefficient (as any decision rule based on the first coefficient is unlikely to generalise well when moving to the test set) and rather compress the original measurements into the 2nd, 3rd, ..., $(D+1)$-th \emph{BAGIDIS} coefficients. $D$ is chosen so as to maximise the cross-validated classification accuracy. Fig. 6 shows how the reconstructions of a randomly chosen supernova become more accurate as $D$ increases.
\begin{figure}
\includegraphics[width=8.5cm]{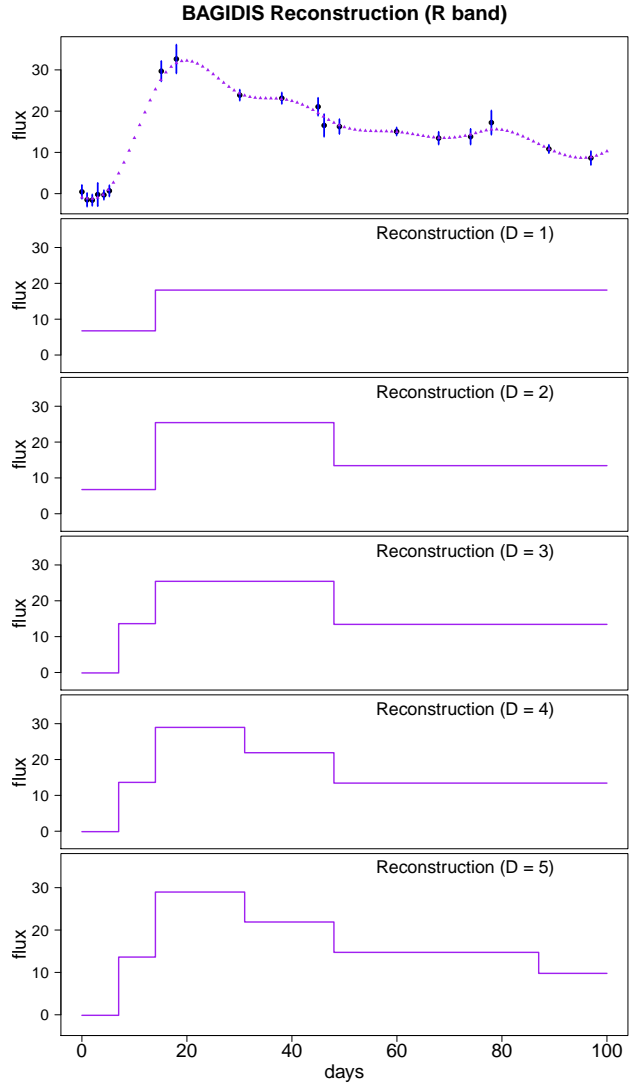}
\caption{The top panel shows the R-band measurements for a randomly chosen supernova (black dots with error bars) together with the spline estimates on the equidistant grid (purple triangles). The next five panels show progressively more accurate reconstructions of the original spline estimates using incrementally more \emph{BAGIDIS} coefficients.}
\end{figure}
Using $D=5$ coefficients in each of the four colour bands provides over 10-fold compression of the spline estimates whilst also preserving the bulk of the light curve information. Using t-distributed stochastic neighbourhood embedding (see \citet{vanderMaaten2008} for more details), we are able to represent the resulting 20-dimensional \emph{BAGIDIS} input vector for each supernova within the combined training and test sets using a two-dimensional subspace (see Fig. 7). Reassuringly, we find that the Ia's and non-Ia's are very well separated in this two-dimensional subspace. Hence, the \emph{BAGIDIS} coefficients are providing excellent discriminatory information for classification.
\begin{figure}
\includegraphics[width=8.5cm]{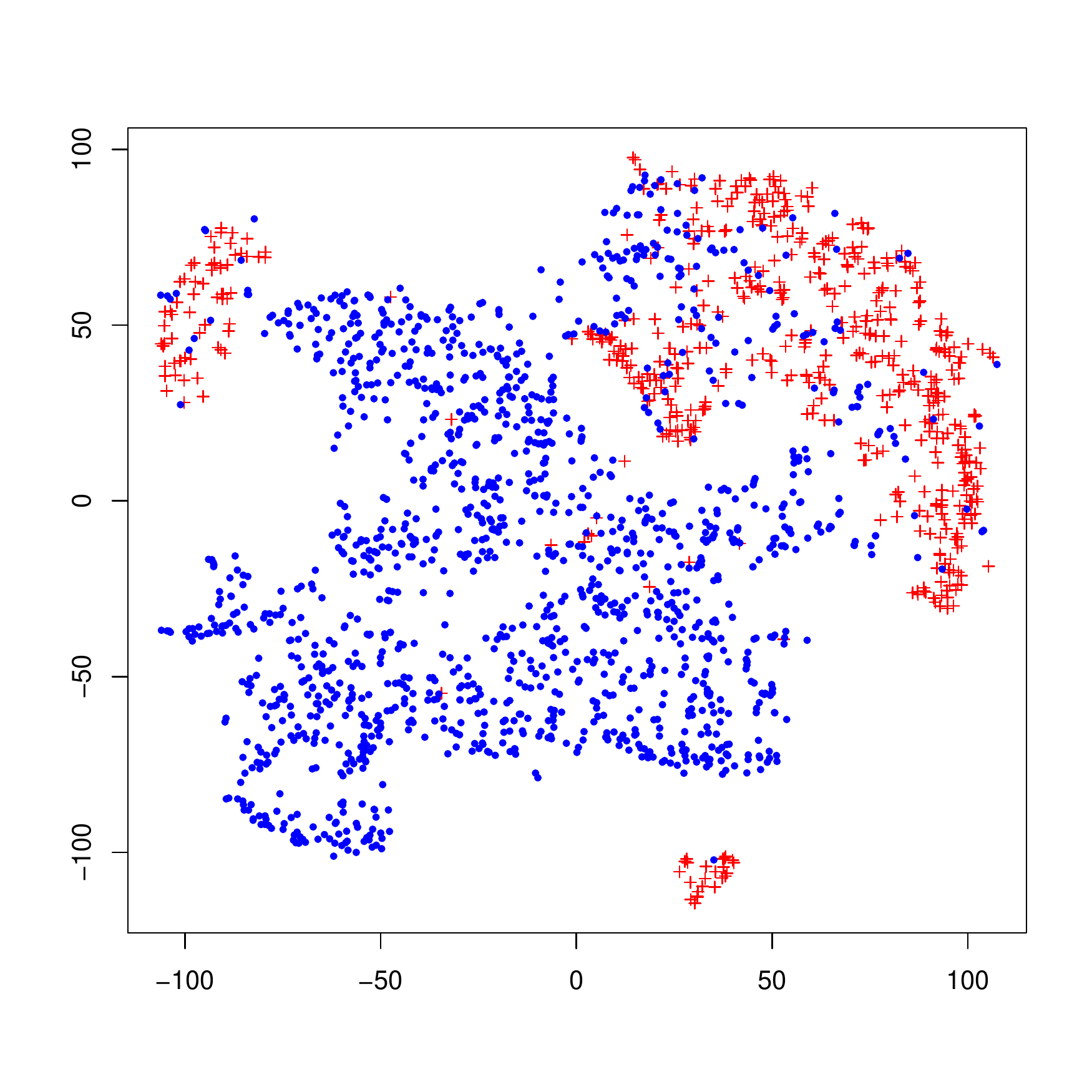}
\caption{A projection of the 20-dimensional \emph{BAGIDIS} coefficients onto a 2-dimensional space using t-distributed Stochastic Neighbourhood Embedding (tSNE). The red pluses correspond to the Ia supernova and the blue points correspond to the non-Ia supernova for 2,000 randomly selected supernovae within the aggregated training and test sets.}
\end{figure}

\begin{figure}
\includegraphics[width=8.5cm]{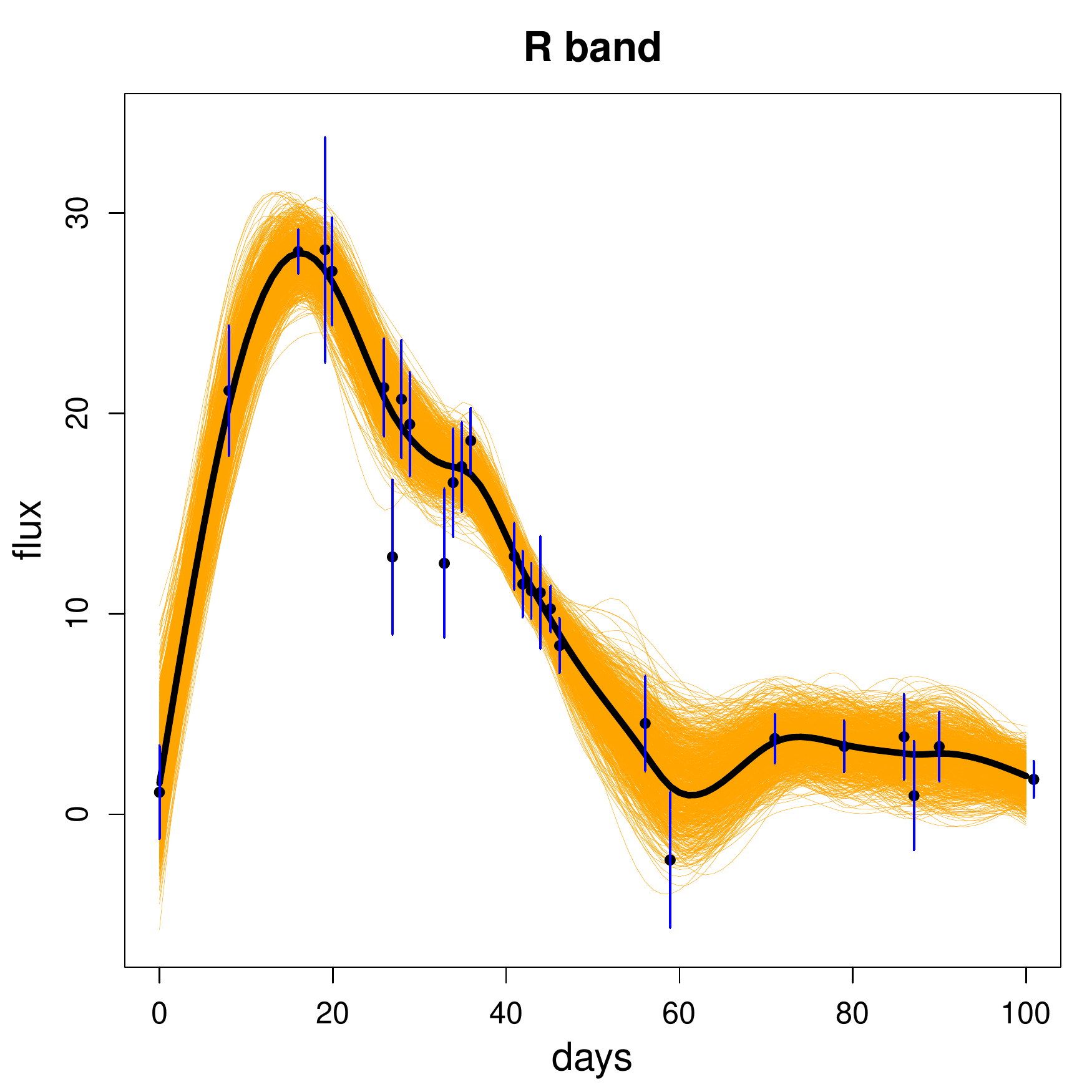}
\caption{The points and error bars represent the R-band flux measurements for a randomly selected supernova. We simulated 1,000 sets of measurements and fitted a spline through each set. These splines are represented by the purple lines. The thick, black line is the spline fit through the original set of measurements.}
\label{splines1000} 
\end{figure}

For each supernova, we thus have four vectors of $D$ \emph{BAGIDIS} coefficients. We expect the randomness in the measurement process to have fed through to the coefficients and so it is important to also calculate a vector of the standard deviations of the \emph{BAGIDIS} coefficients as these may be used to aid in classification. These vectors may be calculated using Monte Carlo simulation (see section 2.2). Fig. \ref{splines1000} shows a randomly selected supernova in the R-band together with the 1,000 splines fits to the simulated measurements. The standard deviations for the R-band were used within the classifiers as cross-validation revealed that their use results in the fewest misclassifications. With both the \emph{BAGIDIS} coefficients and their corresponding standard deviations, we can turn our attention to the classification of each supernova within the test set.

\subsection{Application of Selected Classifiers to the Extracted \emph{BAGIDIS} Coefficients}
We apply three classification algorithms to the SNPCC dataset. Two are well-known classification algorithms and the third is the ranked probability classifier proposed in section 3. For each of these classifiers, we need to estimate the classification performance. \citet{Kessler2010} suggested the following score for the SNPCC:
\begin{equation}
\mathcal{S}_{\mbox{FoM-Ia}}=\frac{N_{\mbox{Ia}}^{\mbox{true}}}{\mathcal{N}_{\mbox{Ia}}^{\mbox{TOT}}}\times\frac{N_{\mbox{Ia}}^{\mbox{true}}}{N_{\mbox{Ia}}^{\mbox{true}}+\mathcal{W}_{\mbox{Ia}}^{\mbox{false}}N_{\mbox{Ia}}^{\mbox{false}}}
\label{score}
\end{equation}
where $N_{\mbox{Ia}}^{\mbox{true}}$ is the number of test set supernovae that are correctly classified as Ia's, $N_{\mbox{Ia}}^{\mbox{false}}$ is the number of test set supernovae that are incorrectly classified as Ia's, $\mathcal{N}_{\mbox{Ia}}^{\mbox{TOT}}$ is the total number of Ia's within the test set and $\mathcal{W}_{\mbox{Ia}}^{\mbox{false}}$ is the false-Ia tag penalty factor. The first fraction represents the \emph{Ia efficiency} -- the proportion of Ia's within the test set that are correctly classified. Also if $\mathcal{W}_{\mbox{Ia}}^{\mbox{false}}$ is set at one, then the second fraction represents the \emph{Ia purity} -- the proportion of test supernovae that we classify as Ia's that are actually Ia's. \citet{Kessler2010} suggested that $\mathcal{W}_{\mbox{Ia}}^{\mbox{false}}$ be set at 3. This creates an extra penalty for incorrectly classifying a non-Ia as a Ia. With $W=3$, the second fraction then represents the \emph{Ia pseudo-purity}. We use this suggested value throughout to evaluate the performance of the three classifiers. 

Since the proportion of Ia's is $\sim 70\%$ within the training set and the Ia proportion is $\sim 30\%$ within the test set, the cross validation estimate of eqn \ref{score} will be biased. We instead use a balanced classification rate to estimate the score. This entails using cross-validation to estimate the \emph{Ia efficiency} and the \emph{non-Ia efficiency} (denoted $\mathcal{E}_{\mbox{Ia}}$ and $\mathcal{E}_{\mbox{non-Ia}}$ respectively). Consequently, the cross-validation estimate using a balanced classification rate will be:
\begin{footnotesize}
\begin{multline}
\hat{\mathcal{S}}_{\mbox{FoM-Ia}}=\mathcal{E}_{\mbox{Ia}}\times\nonumber\\
\left(\frac{L\times p_{\mbox{Ia}}\times\mathcal{E}_{\mbox{Ia}}}{(L\times p_{\mbox{Ia}}\times\mathcal{E}_{\mbox{Ia}})+(\mathcal{W}_{\mbox{Ia}}^{\mbox{false}}\times(L\times (1-p_{\mbox{Ia}})\times(1-\mathcal{E}_{\mbox{non-Ia}})))}\right)
\end{multline}
\end{footnotesize}
where $L$ is the total number of supernova in the test set and $p_{\mbox{Ia}}$ is the expected proportion of Ia's in the test set. In calculating the above score, we set $p_{\mbox{Ia}}=0.3$. Given the above means of predicting the classifier performance, we now turn to the implementation of the three classifiers.
\subsubsection{Nearest Neighbours (NN)}
The Nearest Neighbours (NN) classifier simply compares the distances between the candidate supernovae and the pool of known Ia's and non-Ia's in the \emph{BAGIDIS} coefficient space. The candidate is then classified in the same class as its closest matching supernova within the training set. The distance between the $i$-th and $j$-th supernovae can easily be modified to account for the heteroskedasticity within the dataset. Let:
\[
d(\mathbf{d}_i,\mathbf{d}_j)=\sum_{k=1}^D{\frac{\left|d_{i,k}-d_{j,k}\right|}{\sqrt{{\sigma_{k}^{(i)}}^2+{\sigma_{k}^{(j)}}^2}}}
\]
where $\mathbf{d}_i$ is the $D$-dimensional input vector that contain the relevant \emph{BAGIDIS} coefficients for the $i$-th supernova and ${\sigma_{k}^{(i)}}^2$ is the variance of the $k$-th \emph{BAGIDIS} coefficient for the $i$-th supernova -- as estimated from the Monte Carlo study described in sections 2.2 and 2.3. This distance measure can be calculated for all pairs of supernova that straddle the test and training sets. The only tuning parameter is the number of \emph{BAGIDIS} coefficients $D$ used to calculate the distance. This can be estimated using the balanced classification rates.



\subsubsection{Support Vector Machines (SVMs)}
Support Vector Machines (SVMs) are non-parametric classifiers that are known to be very competitive for many classification problems. They can also be relatively easily modified to account for the heteroskedacity of the data. In addition, SVMs tend to be less susceptible to the curse of dimensionality than NN classifiers. We thus also applied SVM to the SNPCC dataset. SVMs select a classification boundary that maximizes the margin between the boundary and the margin support vectors (selected observations with the dataset). For more details on SVMs see \citet{Hastie2009}. SVMs allow for non-linear decision boundaries via the use of kernels. Kernels are a measure of similarity between two points within the input space. Amongst the most popular kernels is the Radial basis kernel. Here:
\[
K(\mathbf{d},\mathbf{d}')=\exp\left(-\gamma||\mathbf{d}-\mathbf{d}'||^2\right)
\]
where $\gamma$ is a tuning parameter. The radial basis function implicitly assumes that the data is homoskedastic. We thus modify the kernel to account for the SNPCC dataset's heteroskedasticity.
\[
K(\mathbf{d}_i,\mathbf{d}_j)=\exp\left\{-\gamma\sum_{k=1}^D{\left(\frac{d_{i,k}-d_{j,k}}{\sqrt{{\sigma_{k}^{(i)}}^2+{\sigma_{k}^{(j)}}^2}}\right)^2}\right\}
\]
where $\mathbf{d}_i=(d_{i,1},...,d_{i,D})^T$ is a $D$-dimensional input vector that contain the relevant BAGIDIS coefficients. Many SVM packages can accept pre-computed kernels as arguments, which makes it relatively straightforward to run an SVM using the above non-standard kernel.

When running the classifier, there are three tuning parameters whose values may be selected using the cross-validated balanced classification rates: $D$ (the number of \emph{BAGIDIS} coefficients to use as input within the SVM), $\gamma$ (the kernel parameter) and $C$ (a regularisation parameter that determines how tolerant the SVM classification boundary is to misclassification errors within the training set). The larger $C$ is, the less tolerant we are of misclassification errors but the more likely we are to overfit the data. We run a grid search to choose the optimum values for $\gamma$ and $C$ on a logarithmic scale. We use cross-validation to predict the score at the parameter values: $C=2^{-5}, 2^{-3},...,2^{15}$ and $\gamma=2^{-15},2^{-13},...,2^3$. An exponential grid is often used to choose the parameters \citep{vanGestel2004} as $C$ and $\gamma$ are scale parameters. After running the grid search on a very large range of values, we can then narrow in on the sub-space where the classifier is performing best.

\subsubsection{Ranked Probability Classifier}
The ranked probability classifier is simple to implement. For each supernova within the test set, we calculate its probabilistic similarity to each of the training set supernova. 
Due to the false tag penalty within eqn \ref{score}, we may choose to only classify a supernovae with \emph{BAGIDIS} coefficient vector $\mathbf{d}_{\mathcal{C}}$ as a Ia if:
\[
\log p(\mathbf{d}_{\mathcal{C}}|\mathbf{d}_{\mbox{Ia}}^*)>\log p(\mathbf{d}_{\mathcal{C}}|\mathbf{d}_{\mbox{non-Ia}}^*)+V
\]
where $\mathbf{d}_{\mbox{Ia}}^*$ and $\mathbf{d}_{\mbox{non-Ia}}^*$ are the closest matching Ia and non-Ia supernovae respectively and $V$ is a tuning parameter. 
The Ranked Probability classifier would then have three tuning parameters: $D$ (the number of \emph{BAGIDIS} coefficients), $K$ (the number of closest matches against which the candidate object is compared) and $V$. These tuning parameters may be chosen by cross-validation. 

Due to the non-representativity of the training set -- in particular the much higher proportion of Ia within the training set -- the balanced classification rate should be used to calculate the cross-validated score. In many situations, a natural choice would be to set $V=0$ (i.e. classify supernovae to the class with the highest probability) in order to reduce the number of tuning parameters. Cross-validation revealed that the performance of the classifier is roughly constant for $D=2,3,4,5$. In addition, the classifier's performance when $K=1$ was much stronger than when $K=3$ or $K=5$. We thus only present the test scores for the case where $K=1$.

\section{Results for the Classifiers}
\subsection{Results with the true, non-representative training set}
We attempt to classify all test supernovae that have time spans greater than 100 days without using the redshift information. 54\% of supernovae within the SNPCC test dataset satisfy this requirement. In current as well as future photometric surveys, we expect the vast majority of the supernovae captured to satisfy this requirement.

Table \ref{NN} shows the results for the NN classifier when training on various values of $D$. We find that the optimum number of coefficients to train on is two. The two coefficient NN classifier also gave the best cross-validated score. Though only 57\% of the Ia supernovae are identified, the objects that are classified as Ia's have a very high purity ($>90\%$). Table \ref{SVM} shows the the results for the SVM classifier. The SVM tuning parameters $\gamma$ and $C$ were selected by cross-validation on the initial log scale before performing a finer grid search in the best performing region. This finer grid search was also done by cross-validation for the $\gamma$ parameter but did not help in selecting the optimal $C$ on a finer scale. The best performing score of 0.413 is misleadingly optimistic. If $C$ is decreased or increased from 0.727 by even 0.001, the score drops to roughly $\sim 0.300$. The SVM classifier performs worse than the NN classifier. This is perhaps due to its sensitivity to the non-representativity of the training set. Indeed, since for $D=3$ the SVM's classification performance is very sensitive to the value of $C$, it appears that there is a high degree of variability in the SVM's classification rule. 

Table \ref{ProbabilityClassifier} shows the performance of the Ranked Probability Classifier proposed in section 3. The Ranked Probability Classifier is the best performing classifier for all values of $D$. The score increases very slowly as $D$ increases suggesting, for the SNPCC dataset, that $D$ only weakly influences the test score. One of the difficulties in implementing the classifier is that it provides excellent classification of the training set for a wide range of values of the tuning parameter $V$. Table \ref{ProbabilityClassifier} shows its performance for both when $V=0$ (i.e. we classify according to which class has the higher probability) as well as displaying the value of $V$ for which the score on the test set is at its highest. The score increases when we make it easier to classify a point as a Ia -- as can be seen in the fact that the optimum $V$ for all values of $D$ is negative. Reassuringly however, the score does not drop by much if we just use $V=0$ as our default. Hence the Ranked Probability Classifier is relatively robust to the use of a non-optimal value for $V$. In the next section, we compare the relative performances of the three classifiers on a representative training set. 

\begin{table}
\centering
\linespread{1.5}
\caption{Classification results for the Nearest Neighbours (NN) classifier for different number of \emph{BAGIDIS} coefficients. The highest score is shown in bold.}
\begin{tabular}{c c c c c}
\hline
$D$ & Efficiency & Purity & Ps. Purity & Score \\
\hline
\bf{2} & \bf{57.4\%} & \bf{91.2\%} & \bf{77.6\%} & \bf{0.445} \\
3 & 53.1\% & 92.2\% & 79.8\% & 0.424 \\
4 & 57.4\% & 87.1\% & 69.3\% & 0.398 \\
5 & 58.8\% & 86.3\% & 67.8\% & 0.398 \\
\hline
\end{tabular}
\label{NN}
\end{table}

\begin{table}
\centering
\linespread{1.5}
\caption{Classification results for the Support Vector Machine (SVM) classifier for different number of \emph{BAGIDIS} coefficients. The corresponding $\gamma$ and $C$ have been selected at the log-scale using the cross-validated balanced classification rates. The highest score is shown in bold.}
\begin{tabular}{c c c c c c c}
\hline
$D$ & $\gamma$ & $C$ & Efficiency & Purity & Ps. Purity & Score \\
\hline
2 & 0.009 & 0.5&75.3\%&64.0\%&37.2\%&0.280\\
2 & 0.009 & 0.483 & 85.7\% & 61.8\% & 35.0\% & 0.300\\
3 & 0.010 & 0 & 59.6\% & 60.2\% & 33.5\% & 0.200\\
\bf{3} & \bf{0.010} & \bf{0.727} & \bf{68.4\%} & \bf{82.0\%} & \bf{60.3\%} & \bf{0.413}\\
4 & 0.018 & 0 & 61.5\% & 67.4\% & 40.8\% & 0.251\\
4 & 0.018 & 1.006 & 69.9\% & 69.2\% & 42.8\% & 0.299\\
5 & 0.014 & 0 & 62.4\% & 64.9\% & 38.2\% & 0.238\\
5 & 0.014 & 0.910 & 71.2\% & 68.0\% & 41.4\% & 0.295\\
\hline
\end{tabular}
\label{SVM}
\end{table}

\begin{table}
\centering
\linespread{1.5}
\caption{Classification results for the Ranked Probability classifier for different number of \emph{BAGIDIS} coefficients. $V$ is a tuning parameter (see section 5.2.3). The highest score is shown in bold.}
\begin{tabular}{c c c c c c}
\hline
$D$ &$V$& Efficiency & Purity & Ps. Purity & Score \\
\hline
2 & 0 & 72.6\% & 84.5\% & 64.4\% & 0.468\\
2 & -1.1 & 75.0\% & 83.7\% & 63.1\% & 0.473\\
3 & 0 & 75.2\% & 83.6\% & 63.0\% & 0.474\\
3 & -2.3 & 79.7\% & 81.7\% & 59.8\% & 0.477\\
4 & 0 & 75.7\% & 83.0\% & 62.0\% & 0.470\\
4 & -1.3 & 78.6\% & 82.6\% & 61.2\% & 0.481\\
5 & 0 & 77.5\% & 83.1\% & 62.1\% & 0.481\\
\bf{5} & \bf{-1.4} & \bf{80.5\%} & \bf{82.4\%} & \bf{60.9\%} & \bf{0.490}\\
\hline
\end{tabular}
\label{ProbabilityClassifier}
\end{table}

\subsection{Results with a representative training set}
We now turn our attention to the performance of the classifiers if we were able to run the classifiers on a representative training set. Since it is unrealistic to have a truly representative training set, we wish to discern the robustness of the three classifiers to the non-representativity of the training set. Also, if classification algorithms are egregiously affected by the non-representativity, it would motivate the investment of substantial resources towards the production of a more representative training set.

Of the roughly 10,000 supernovae with a timespan of at least 100 days, less than 5\% have known class. The organisers of the challenge later released the classes of all supernovae, so we can randomly assign roughly 5\% of supernovae to have a known class and the remaining 95\%  to be part of the test set. Subsequently we train all three classifiers on this new training set. Recall that we previously avoided using the first \emph{BAGIDIS} coefficient (which stores the mean flux measurement) for classification since any classification rule based on the information within the first coefficient would be overly sensitive to the training supernovae being brighter on average than the test supernovae. Now that we are training on a representative training set, including the information within the first coefficient will improve performance. However in the interest of being able to directly compare the results with the classifiers that use the true training set, we still choose to avoid using the first coefficient in all three classifiers.

\begin{table}
\centering
\linespread{1.5}
\caption{Classification results for the Nearest Neighbours (NN) classifier on a representative training set for different number of \emph{BAGIDIS} coefficients. The highest score is shown in bold.}
\begin{tabular}{c c c c c}
\hline
$D$ & Efficiency & Purity & Ps. Purity & Score \\
\hline
2 & 70.7\% & 84.2\% & 64.0\% & 0.452 \\
3 & 76.7\% & 84.9\% & 65.2\% & 0.500 \\
\bf{4} & \bf{76.4\%} & \bf{86.2\%} & \bf{67.6\%} & \bf{0.517} \\
5 & 78.1\% & 85.3\% & 65.8\% & 0.514 \\
\hline
\end{tabular}
\label{NNrep}
\end{table}

\begin{table}
\centering
\linespread{1.5}
\caption{Classification results for the Support Vector Machine (SVM) classifier on a representative training set for different number of \emph{BAGIDIS} coefficients. The corresponding $\gamma$ and $C$ have been selected using the cross-validated balanced classification rates. The highest score is shown in bold.}
\begin{tabular}{c c c c c c c}
\hline
$D$ & $\gamma$ & $C$ & Efficiency & Purity & Ps. Purity & Score \\
\hline
2 & 0.006 & 0.031 &71.5\%&80.7\%&58.3\%&0.417\\
2 & 0.006 & 0.043 & 76.6\% & 78.7\% & 55.1\% & 0.423\\
3 & 0.016 & 0.125 & 77.2\% & 73.8\% & 48.5\% & 0.374\\
3 & 0.016 & 0.109 & 78.1\% & 75.3\% & 50.5\% & 0.394\\
4 & 0.015 & 0.125 & 77.5\% & 73.1\% & 47.6\% & 0.369\\
4 & 0.015 & 0.106 & 78.6\% & 75.7\% & 50.9\% & 0.400\\
5 & 0.005 & 0.031 & 71.4\% & 81.2\% & 59.0\% & 0.421\\
\bf{5} & \bf{0.005} & \bf{0.036} & \bf{74.9\%} & \bf{80.2\%} & \bf{57.4\%} & \bf{0.430}\\
\hline
\end{tabular}
\label{SVMrep}
\end{table}

\begin{table}
\centering
\linespread{1.5}
\caption{Classification results for the Ranked Probability classifier on a representative training set for different number of \emph{BAGIDIS} coefficients. The highest score is shown in bold.}
\begin{tabular}{c c c c c c}
\hline
$D$ &$V$& Efficiency & Purity & Ps. Purity & Score \\
\hline
2 & 0 & 88.2\% & 83.0\% & 61.9\% & 0.546\\
\bf{2} & \bf{0.4} & \bf{85.7\%} & \bf{84.3\%} & \bf{64.2\%} & \bf{0.550}\\
3 & 0 & 88.7\% & 82.5\% & 61.1\% & 0.542\\
3 & -0.1 & 89.5\% & 82.2\% & 60.6\% & 0.543\\
4 & 0 & 89.2\% & 81.8\% & 60.0\% & 0.535\\
4 & 0.7 & 85.9\% & 83.6\% & 62.9\% & 0.541\\
5 & 0 & 89.1\% & 81.4\% & 59.3\% & 0.528\\
5 & 1.1 & 84.9\% & 83.9\% & 63.4\% & 0.539\\
\hline
\end{tabular}
\label{ProbabilityClassifierrep}
\end{table}

Tables \ref{NNrep}, \ref{SVMrep} and \ref{ProbabilityClassifierrep} show the the performances on a representative training set for the NN, SVM and Ranked Probability classifiers respectively. As expected, the performance of all three classifiers improves when using a representative training set. However, even with a representative training set, the SVM still is the worst performing classifier and the Ranked Probability classifier is still the top performing classifier. We also wish to compare the relative robustness of the three classifiers to the non-representativity of the training set. We thus look at the percentage increase in the score of each classifier when using a representative dataset. Note for the SVM classifier, we are able to reliably choose the value of the tuning parameter $C$ at the log-scale but not on a finer scale. Hence in calculating the percentage increase for the SVM we only consider the scores where $C$ is set equal to $2^{(2\times j)+1}$ and $j$ is an integer. Likewise, since for the Ranked Probability classifier, the classification on the training set is excellent for a wide range of $V$ values (thus preventing us from choosing $V$ by cross-validation), we only calculate the percentage increase for the cases where $V=0$. Based on these restrictions, the average percentage increase in the performance of the NN, SVM and Ranked Probability classifiers is 19.6\%, 65.0\% and 13.7\% respectively. We thus see that the NN and Ranked Probability classifiers are more robust to the non-representativity of the training set. This robustness to non-representativity is probably due to both classifiers ultimately choosing the class of a candidate supernovae based only on the closest matching Ia and non-Ia supernovae within the training set. The subset of the training set that is composed of supernovae that are the closest matching Ia or non-Ia to at least one of the candidate supernovae are by definition more representative of the test set.

\section[]{Conclusion}
We propose a generalized method for classifying astronomical objects based on functional data, which includes light curves and astronomical spectra. Classification based on functional data has a number of obstacles. The datasets are often not standardized and the data for any one astronomical object will often exist in a very high dimensional space. High dimensional spaces are problematic as the algorithms take more computational resources and the variance of the resulting classification boundary often increases exponentially with the dimensionality of the space.  Our proposal fits splines to the functional data and samples the spline estimates on a regularly spaced grid as a means to standardize the datasets. After the datasets have been standardized, the \emph{BAGIDIS} wavelet methodology was demonstrated to be an effective means to reduce the dimensionality of the standardized functional data whilst also minimizing information loss. The resulting vector of \emph{BAGIDIS} coefficients can be used instead of the original functional data as features to input to any classifier with the advantage that the same set of wavelet coefficients are produced for any curves that are identical apart from a shift in location. This property is highly advantageous as we don't have to align curves before applying the procedure. Alignment is often a computationally expensive procedure that necessitates some underlying assumptions about the behaviour of the functional data and introduces extra noise into the classification. Hence, the \emph{BAGIDIS} methodology allows for a genuinely non-parametric approach to data compression.

Astronomical training sets often tend to be at lower redshifts and have lower (that is brighter) apparent magnitudes than the test set. Unfortunately, the robustness of classifiers to non-representative training sets is an under-researched area. We believe that classifiers that base the classification rule for a candidate object on the closest matching training object in each of the classes will tend to be more robust to the non-representativity of the input space since training data that are the closest matches to the test set should be more representative of the test set. The nearest neighbours classifier (NN) is an example of such a classifier. We also propose the Ranked Probability classifier. This classifier uses a probabilistic measure of similarity to determine the closest matching training object in each class.

We applied the proposed methodology to the Supernova Photometric Classification Challenge (SNPCC). To the best of our knowledge, this is the first analysis that does not require the curves to be re-aligned so that peak brightness is at time zero. Such a re-alignment can be computationally expensive and makes assumptions about the form of the supernovae light curves. The challenge consisted of trying to identify the Ia supernova based on a series of flux measurements. The quality of this functional data varied drastically between the supernovae; both in terms of the number of flux measurements and the measurement error bars. Despite this, we were able to standardize the dataset by fitting a spline to the data and taking the resulting spline estimates on an equidistant grid. The data within this series of $n$ spline estimates was compressed into $D$ \emph{BAGIDIS} coefficients ($D \ll n$). A t-SNE plot revealed that the second to the sixth \emph{BAGIDIS} coefficients provided excellent discriminatory information on the class of the supernova.  We then applied an NN, an SVM and the Ranked Probability classifier to all those supernovae with a timespan greater than 100 days. This constituted 54\% of the SNPCC test dataset and we expect the percentage to be much larger in future photometric surveys. Both the NN and the Ranked Probability classifier provided highly competitive results with the Ranked Probability classifier offering over 80\% \emph{Ia efficiency} and \emph{purity}. We also noted that both classifiers appeared to be more robust to the non-representativity of the training set than the SVM classifier. We speculate that this could be due to both classifiers basing the classification rule on a subset of the training set that is more representative of the test set. Though the methodology presented in this paper was applied to light curve data, it can also be readily applied to other types of astronomical functional data such as spectra. The \emph{BAGIDIS} methodology could potentially also be used to summarise the light curves captured by future large synoptic sky surveys.

\section*{Acknowedgements}
We thank Benoit Frenay from the University of Namur for helpful discussions. Melvin Varughese was funded by the National Research Foundation of South Africa as well as a Research Development Grant from the University of Cape Town. Rainer von Sachs gratefully acknowledges funding by contract Projet d'Actions de Recherche Concert\'eesÓ No. 12/17-045 of the Communaut\'e fran\c{c}aise de BelgiqueÓ and by IAP research network Grant P7/06 of the Belgian government (Belgian Science Policy). He also thanks the Belgian FNRS for supporting his visit to UCT in spring 2013, where this research project took its origins. Bruce Bassett is supported by the National Research Foundation of South Africa. We thank the referee for his/her constructive comments that have helped improve this paper.

\label{lastpage}
\bibliographystyle{mn2e}
\bibliography{references}
\end{document}